\documentclass[namedreferences]{SolarPhysics}
\usepackage[optionalrh]{spr-sola-addons} 
\usepackage{graphicx}        
\usepackage{color}           
\usepackage{url}             

\begin{document}

\begin{article}

\begin{opening}

\title{Acoustic Power Absorption  and its Relation with Vector Magnetic Field of a Sunspot }

\author{S.~\surname{Gosain}\sep
        S.~K.~\surname{Mathew}\sep
        P.~\surname{Venkatakrishnan}
       }
\runningauthor{S. Gosain \textit{et al.}}
\runningtitle{Acoustic Power Absorption}
   \institute{ Udaipur Solar Observatory, Post Box 198, Dewali,
   Udaipur--313001, India
                     email: \url{sgosain@prl.res.in} \\
             }

\begin{abstract}
The distribution of acoustic power over sunspots shows an
enhanced absorption near the umbra--penumbra boundary. Earlier
studies revealed that the region of enhanced absorption
coincides with the region of strongest transverse potential
field. The aim of this paper is to \textit{i}) utilize  the
high-resolution vector magnetograms derived using {\it Hinode}
SOT/SP observations and study the relationship between the
vector magnetic field  and power absorption and \textit{ii})
study the variation of power absorption in sunspot penumbrae
due to the presence of  spine-like radial structures.   It is
found that \textit{i}) both potential and observed transverse
fields peak at a similar radial distance from the center of the
sunspot, and \textit{ii})  the magnitude of the transverse
field, derived from {\it Hinode} observations, is much larger
than the potential transverse field derived from SOHO/MDI
longitudinal field observations. In the penumbra, the radial
structures called spines (intra-spines) have stronger (weaker)
field strength and are more vertical (horizontal). The
absorption of acoustic power in the spine and intra-spine shows
different behaviour with the absorption being larger in the
spine as compared to the intra-spine.

\end{abstract}
\end{opening}

\section{Introduction}
     \label{S-Introduction}
In the recent years, helioseismic observations of the Sun,
meant to study the solar interior, have been used extensively
to study the \textit{p}-mode interaction with surface magnetic
fields. It is known  that the \textit{p}-mode amplitude is
suppressed in sunspots and in regions of strong magnetic field
\cite{Leighton1962,Lites1982}. This reduction in
\textit{p}-mode amplitude is attributed to the absorption of
acoustic waves by the magnetic structures \cite{Cally1995}. In
fact, observations show that the amplitude of \textit{p}-mode
oscillations decreases  while the amplitude of high-frequency
oscillations increases with increasing line-of-sight (LOS)
magnetic field, with a transition near the acoustic-cutoff of
5.3 mHz \cite{Hindman1998,pvk2002}. Further, the incoming
acoustic power was seen to be greater than outgoing power,
which implies absorption of acoustic energy by magnetic
structures. About 50\% of the energy of acoustic
\textit{p}-modes is absorbed in the regions of magnetic-field
concentrations like sunspots \cite{Braun1987,Braun1988}.

The  acoustic-power absorption in sunspots could be a valuable
tool for probing the sub-photospheric structure of the sunspot
magnetic field, once the physical processes responsible for the
absorption are understood. Two theories have been proposed to
explain this absorption. The first proposal is that of resonant
absorption based on strong dissipation of  MHD waves in a
sunspot flux tube resulting in a net loss of energy. The
resonant absorption has been  well studied in coronal loops
\cite{Ionson1982,Hollweg1984,Davila1987}. The second proposal
is that of mode conversion, where the incoming acoustic energy
is transferred through mode conversion into magnetic wave-modes
\cite{Hindman1997}. These waves can leak from the system,
either upwards through chromosphere \cite{Bart2004}, or
downward through the base of the sunspot. However, both of
these processes (\textit{i.e.}, the resonant absorption and
mode conversion)  do not exclude each other and may be
operating simultaneously.

The acoustic-power distribution in sunspots shows an enhanced
absorption of power near the umbra--penumbra boundary
\cite{Mathew2008}. Further, this region of enhanced absorption
was shown to coincide with the region where the mean
inclination is about 45 degrees and the transverse field is
maximum \cite{Mathew2008}. The inclination angle and transverse
fields were however obtained by computing the potential field
using line-of-sight field from SOHO/MDI. With the availability
of high-quality vector magnetic-field observations by {\it
Hinode}/SOT, we can now study the relationship between the
vector magnetic field and the distribution of acoustic power in
a sunspot.

 In this paper, we study the relationship between the vector
magnetic field and the acoustic-power absorption in different
structures of a sunspot. The radial, as well as the azimuthal,
variation of magnetic parameters and power absorption is
studied. Also, we compare the observed and potential transverse
field of a sunspot. Section 2 describes the observational data,
its reduction, and analysis. Section 3 describes the results
and in the last section we discuss the results and present
conclusions.

\section{Observations}

In order to compare the vector magnetic-field parameters of a
sunspot and the acoustic signals, we selected an observational
data set which satisfies the following criteria: \textit{i})
The observations must be simultaneous, \textit{ii}) a simple
circular unipolar sunspot should be selected as the azimuthal
symmetry simplifies our analysis, and \textit{iii}) the sunspot
should be located close to disk center for minimal projection
effects.

\subsection{SOHO/MDI observations}

\begin{figure}
\centerline{\includegraphics[width=0.45\textwidth,clip=]{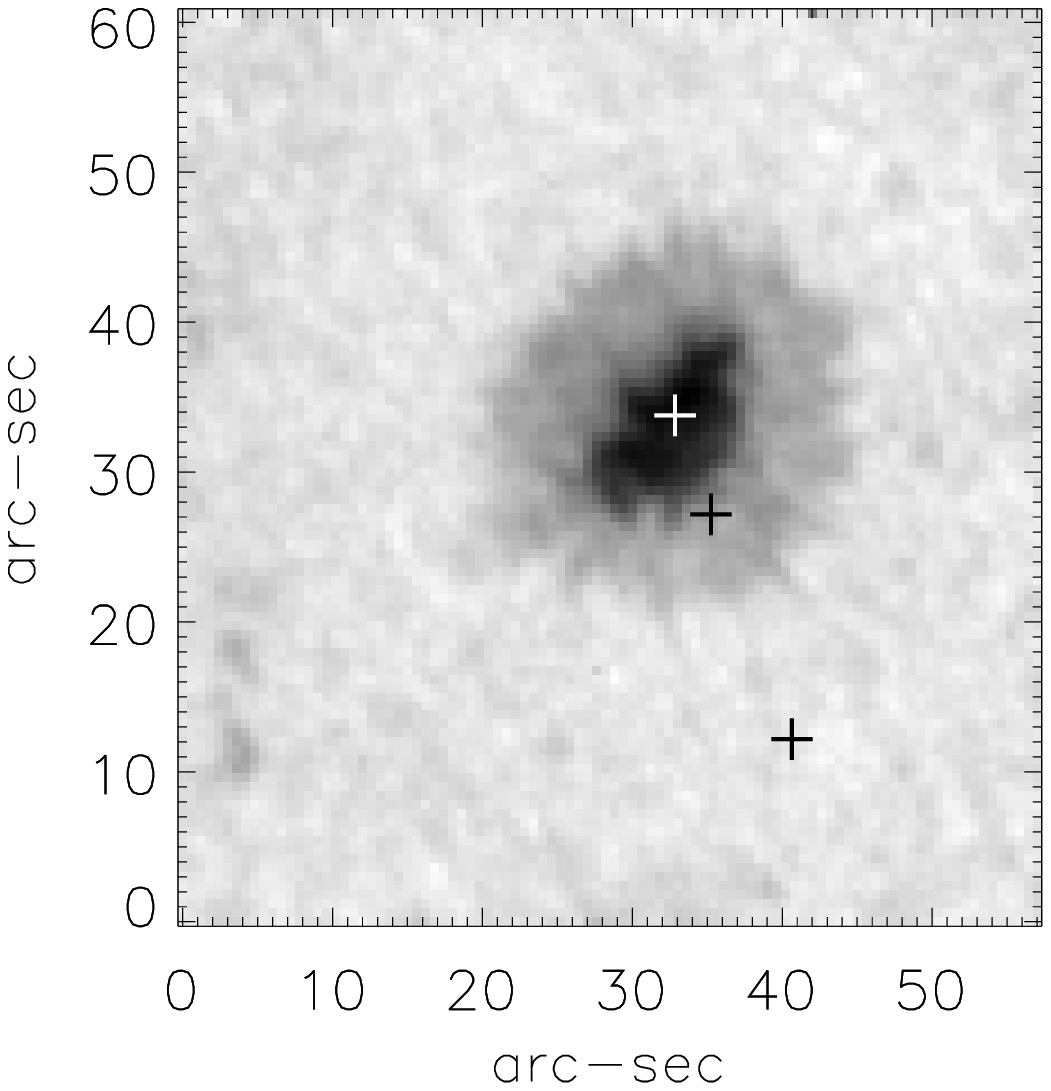}}
\centerline{\includegraphics[width=1.0\textwidth,clip=]{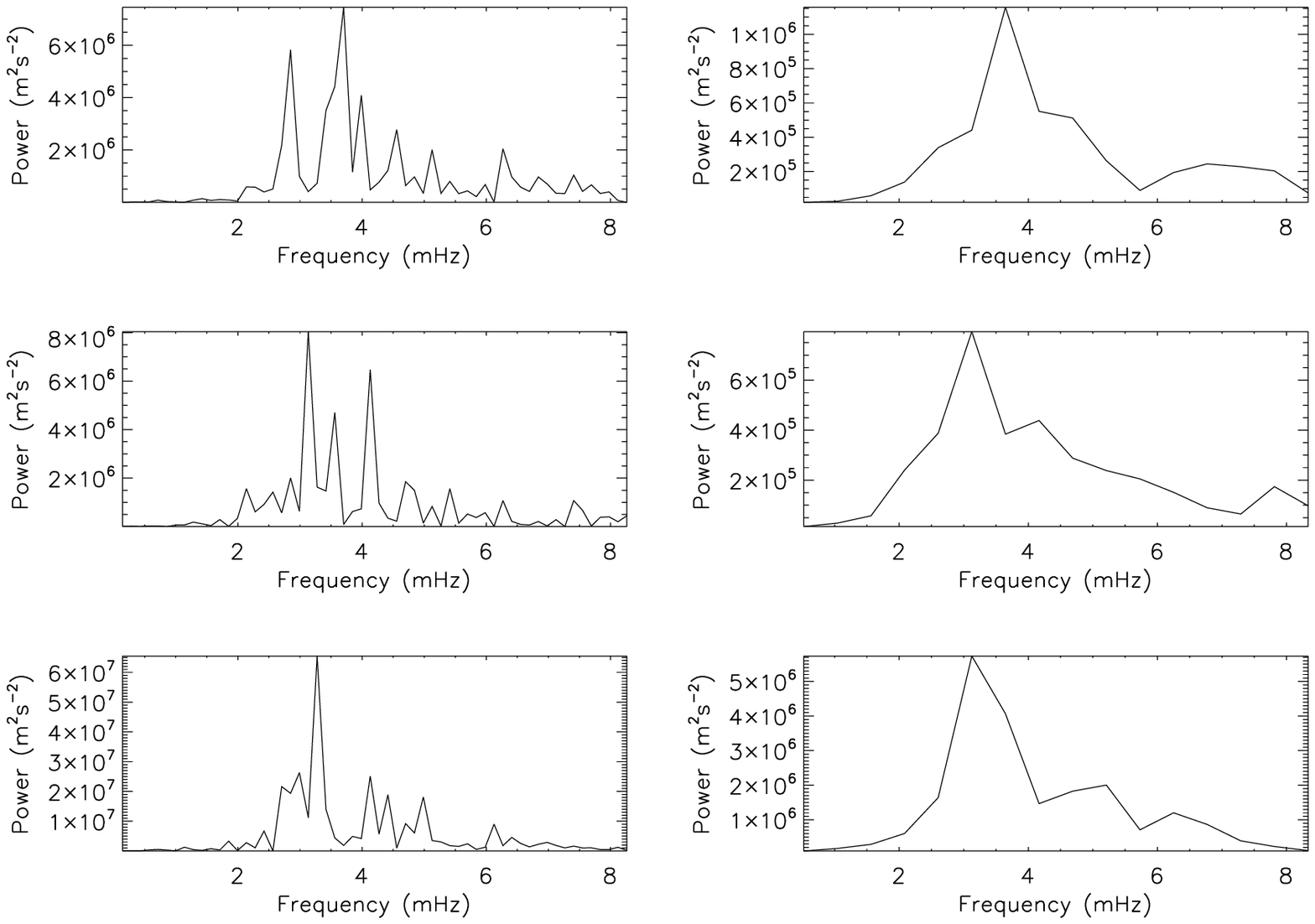}}
\caption{The continuum MDI image of the sunspot with three points marked in umbra, penumbra, and outside the sunspot is shown at the top.
The  power spectra without (left) and with smoothing (right) for the three points marked on the image above. The top, middle
and bottom row correspond to umbra, penumbra, and point outside the sunspot.  }
\label{gosain:fig1}
\end{figure}

The data for a sunspot in NOAA AR 10960 at the disk center
($\mu$=0.98) on 08 June 2007 08:59 -- 10:58 UT met our
criteria. These dopplergrams were obtained by SOHO/MDI using
high-resolution mode with an image scale of  0.6 arc-sec per
pixel. Solar rotation was removed by alignment of images using
image cross-correlation technique. The slowly varying component
of the velocity signal was removed by doing a running
difference \textit{i.e.}, subtracting consecutive dopplergrams.
Thus, a time series of two hours with one-minute cadence, which
corresponds to a frequency resolution of 0.13 mHz, was
constructed. The resulting power spectra at three locations in
the umbra, penumbra and the surrounding moat region is shown in
Figure~\ref{gosain:fig1}. The peak corresponding to the
dominant five minute oscillations can be noticed clearly and
this peak is seen to decrease with increasing field strength in
the umbra. Also, we note that the spectrum is quite noisy (left
panels). To improve the signal-to-noise ratio we reduce the
spectral resolution.

For each pixel we split a single  $N$ point time series
$T_{1}=[1,2,3,...N]$ into $k$ smaller time series of size $n$
such that $t_{1}=[1,2,3,..n]$, $t_{2}=[2,3,..n+1]$, .......
$t_{k}=[N-n,..N]$ and add power spectra for each of the $k$
time series.  This power spectra is then divided by the number
$N$ to obtain an average power spectra. Although this operation
reduces the frequency resolution (0.52 mHz in our case for
$N$=118, $n$=33 and $k$=80), the power spectra becomes less
noisy. The peak of the power spectrum is well defined for the
smoothed spectra as compared to the multiple peaks in the raw
power spectra.  The power maps are constructed by summing the
power in the band 1.6 to 6.6 mHz. The choice of $n$ is made
such that we have sufficient number of sample points in the
spectral band 1.6 to 6.6 mHz for making power map. We arrive at
$n$=33 which leads to about 10 data points in the spectral
band. Choosing $n$ too small would not lead to smooth power
spectra while choosing $n$ too large would not lead to
sufficient number of data points in spectral band 1.6 to 6.6
mHz.

\subsection{{\it Hinode} SOT/SP observations}

The space-based {\it Hinode}/SOT observes active regions with
the onboard  Spectropolarimeter (SP) in the Fe I 630.15 nm and
630.25 nm line pair \cite{Kosugi2007,Lites2007,Tsuneta2008,
Ichimoto2008}. A map of an active region is made by scanning
the slit of the spectrograph.  First, the Stokes profiles were
inverted using the ``HeLiX'' inversion code \cite{Lagg2004} to
generate magnetic maps of the scanned region.  This inversion
code fits the observed Stokes profiles with synthetic ones that
are obtained from an analytic solution of the Unno-Rachkovsky
equations under Milne-Eddington model atmosphere assumptions
\cite{Unno,Rachkovsky,Landolfi}. The model atmosphere used in
our present inversion included a single component atmosphere
with a stray-light component. Further, the azimuthal ambiguity
of transverse-field component was resolved using the acute
angle method \cite{Harvey69}. The magnetic-field vectors were
then transformed from observed to the local solar reference
frame using the procedure described by Venkatakrishnan and Gary
(1989). The region corresponding to the sunspot was extracted
from the MDI continuum images and the intensity map from {\it
Hinode}/SP was re-scaled and registered to match the MDI image.
Before registration, the spatial resolution of {\it Hinode}/SP
continuum images was degraded to the MDI resolution. The fine
registration was done by cross-correlating the two images. The
radial and azimuthal variations of the observed parameters of
the two registered datasets are studied and the results are
presented.

\section{Results }
\subsection{Azimuthal Average of Parameters and their Radial Variation}
\begin{figure}
\centerline{\includegraphics[width=0.45\textwidth,clip=]{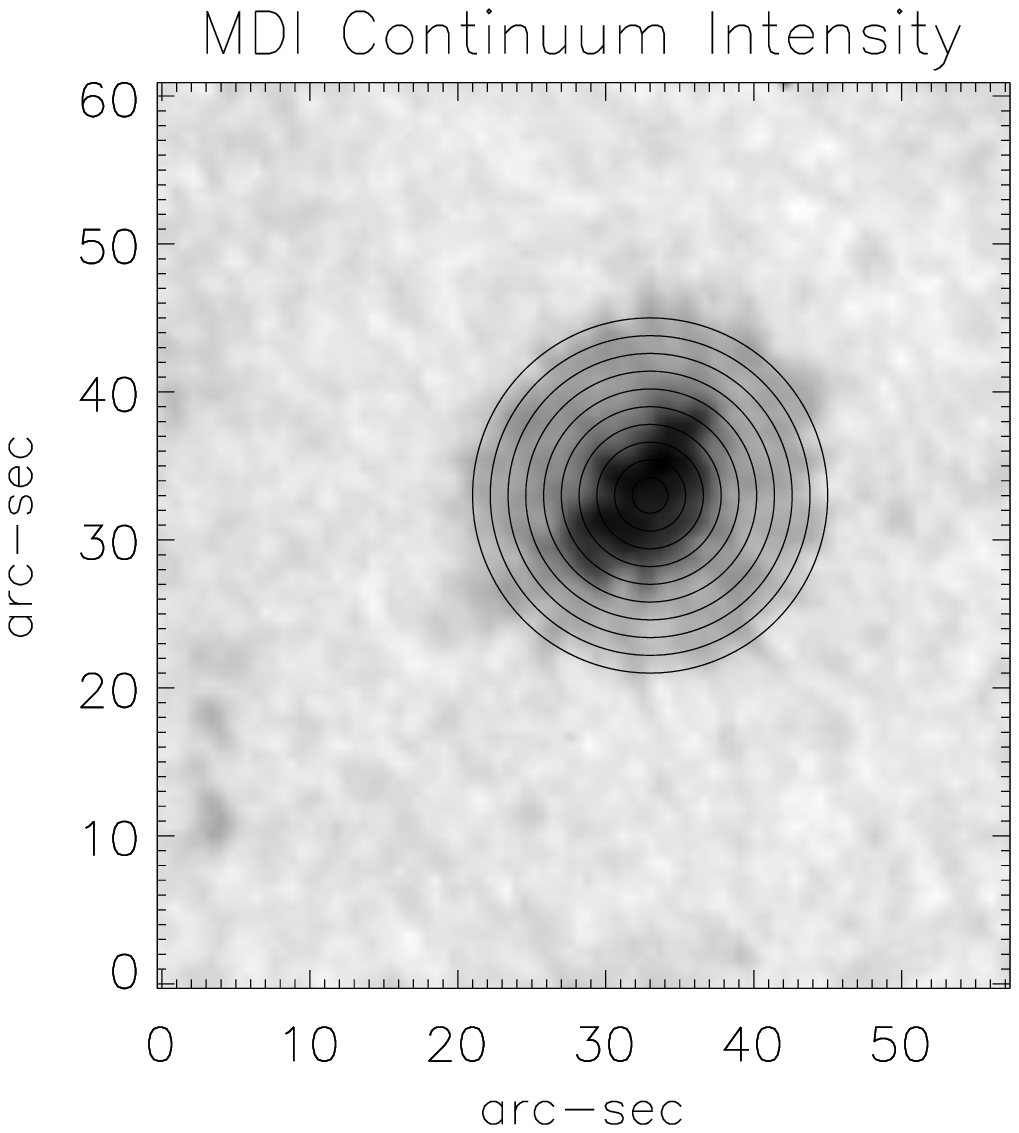}}
\centerline{\includegraphics[width=0.95\textwidth,clip=]{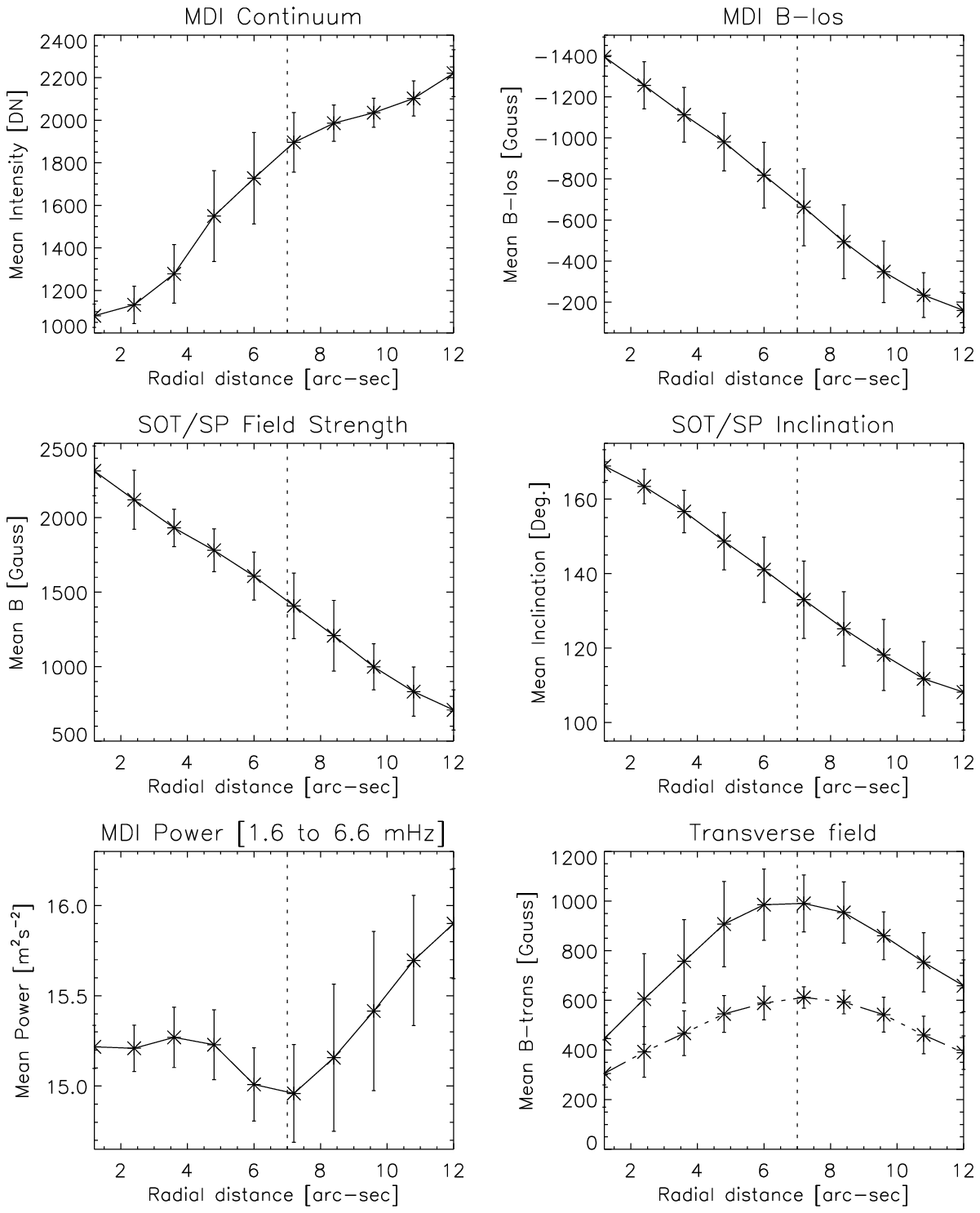}}
\caption{ Concentric rings with increasing radius drawn over sunspot as shown in the upper panel are used for studying the radial
variation of different parameters. The average value of the parameters is calculated along each ring which is plotted in the lower panels.
The radial distance from the center of the sunspot is given in arcsec.
The right panel in the bottom row represents observed transverse field ({\it solid line}) and  potential transverse field computed using
line-of-sight field measured by MDI ({\it dash-dotted line}). The vertical dashed line in all panels marks the location of enhanced power
absorption.}
\label{gosain:fig2}
\end{figure}

The radial variation of the observed parameters is shown in
Figure~\ref{gosain:fig2}. The concentric rings, spaced 1.2
arcsec apart, shown in the top panel, represent the paths along
which the various parameters are averaged. These mean values
are then plotted in the lower panels with increasing radial
distance from the center of the rings. The standard deviation
(1$\sigma$) of these fluctuations is represented by the length
of the vertical bars plotted over the mean radial profiles. The
mean profiles for continuum intensity, line-of-sight field,
field strength and field inclination show a smooth variation
with radial distance. However, the acoustic-power in the 1.6 to
6.6 mHz band shows a discontinuous behaviour at the
umbra--penumbra boundary. As one approaches umbra--penumbra
boundary there is a region of little enhancement followed by
region of reduction in acoustic power.  This effect was also
seen earlier in the study of a different sunspot
\cite{Mathew2008}.

Further, the solid curve in the bottom right panel in
Figure~\ref{gosain:fig2} shows the variation of observed
transverse field with radial distance. The potential transverse
field, calculated using the MDI line-of-sight (LOS) field as
boundary conditions, is shown with dot-dashed curve. The
potential transverse field is significantly underestimated as
compared to the transverse field  observed by SOT/SP due to
different measurement techniques as well as different spatial
and spectral resolutions of the two instruments.  The cross
comparison of the magnetic field measurements by the SOHO/MDI
and SOT/SP instruments shows that the field strength in MDI
observations is underestimated by about 70\% \cite{Wang2009}.
An underestimate of the potential transverse  field derived
from MDI observations as compared to  transverse field derived
from the {\it Hinode} SOT/SP observations, shown in the right
panel of the bottom row in Figure~\ref{gosain:fig2},
approximately follows a similar relation. It may be noted that
the MDI magnetogram observations that are used in the present
work are level 1.8.1 magnetograms which have been corrected for
spatially dependent sensitivity variations as determined by
Tran \textit{et al.} (2005), leading to an increase in flux
values by a factor of about 1.7, as compared to level 1.8.0
magnetograms. More recent (level 1.8.2) magnetograms have been
further corrected for sensitivity variations, determined by
Ulrich \textit{et al.} (2009). According to these new
calibrations the flux values at the disk center reduce by about
8\% and limb values reduce by about 30 \%. The sunspot in the
present study is located close to disk center ($\mu$=0.98). So,
the flux values in the (level 1.8.1) magnetograms used in this
study may change by about 8\%. Further, we note that in a
parallel study Demidov and Balthsar (2009) has suggested that
the re-calibration of MDI magnetograms is perhaps not
straightforward and that more detailed studies are required.
Irrespective of the magnitude of the MDI field, we can see that
both curves in the bottom right panel in
Figure~\ref{gosain:fig2} show the location of enhanced
\textit{p}-mode power absorption and maximum transverse field
occur at a similar radial distance from the sunspot umbra. This
distance is marked by the vertical dashed line in all panels of
Figure~\ref{gosain:fig2}. The inclination of the magnetic field
at this location is also about 135 degrees (or -45 degrees).
This is in agreement with the previous results
\cite{Mathew2008}.

From the profiles of magnetic parameters and acoustic power,
shown  in Figure 2, it can be seen that on the right-hand-side
(RHS) of the vertical dashed line the magnetic parameters, as
well as the acoustic power, follows a smooth variation such
that any of the magnetic parameters can be correlated or
anti-correlated with the the acoustic power. While the profile
on the left-hand-side (LHS) of the vertical dashed line
exhibits no clear relation between the magnetic parameters and
acoustic power. This implies that none of the magnetic
parameters can fully explain the acoustic power absorption in
the entire sunspot. However, to compare the relative importance
of the magnetic parameters for acoustic power absorption in a
sunspot, we do a correlation analysis of these parameters in
the following section.

\subsection{Magnetic Parameters versus Acoustic Power Absorption in Different Regions of Sunspot}
Here we study the relation between different magnetic
parameters and acoustic power absorption in the entire sunspot
as well as in different regions of the sunspot. We isolate
three regions in a sunspot, \textit{i.e.}, umbra,
umbra--penumbra boundary and penumbra, using continuum
intensity thresholds. These regions are shown by the contours
in Figure 3. The scatterplot of the magnetic-field strength,
transverse field, and inclination angle \textit{versus}
acoustic power absorption is shown in Figure 4 for these
regions. The top panel shows the scatterplot for the entire
sunspot area {\it i.e.}, pixels within the outermost contour.
Second, third, and fourth panel from the top, in Figure 4, show
the scatterplot for the umbra, umbra--penumbra boundary and
penumbra, respectively.

 We compute the linear Pearson correlation coefficient (R)
between the acoustic power and the magnetic parameters for
different regions of the sunspot using the relation
$$R=\frac{\frac{1}{N}\sum_{i=1}^{N}(P_i-\bar{P})(M_i-\bar{M})}{\sigma_{P} ~\sigma_{M}}~ \times 100$$
Where, the numerator gives the covariance between the acoustic
power $P$ and one of the magnetic parameters $M$ ({\it i.e.},
field strength, transverse field strength or inclination
angle). While, the denominator is the product of the standard
deviation of $P$ and $M$, given by $\sigma_{P}$ and
$\sigma_{M}$ respectively. This correlation coefficient shows
the goodness of fit for a linear model ({\it i.e.}, a measure
of the dispersion of data points from a linear slope).  For the
whole sunspot, we find that there is an anti-correlation
between the acoustic power and all of the three parameters {\it
i.e.}, the field strength (65\%), the transverse field (61\%)
and the field inclination (34\%). While in the umbra the
scatter is rather flat showing  weaker anti-correlation  of
6\%, 12\%, and 10\% between acoustic power and field strength,
transverse field and field inclination, respectively. For the
umbra--penumbra boundary, which corresponds to the region of
enhanced absorption of \textit{p}-mode power, the acoustic
power is smaller as compared to the umbra and here also the
anti-correlation  is weaker {\it i.e.}, 16\%, 17\% and 15\%
between acoustic power and field strength, transverse field and
field inclination, respectively. While in the penumbra, the
acoustic power is anti-correlated with field strength,
transverse field, and field inclination by 46\%, 50\%, and
13\%, respectively.

Thus, the maximum anti-correlation of acoustic power is with
the field strength in the entire sunspot, while it is with the
transverse field in the penumbral region. The correlation of
acoustic power with the field inclination remains weak in
general.

\subsection{Azimuthal Variation of Parameters along mid-Penumbra}
Figure~\ref{gosain:fig3} shows maps of various observed
parameters of the sunspot: the continuum image,  transverse
field, field strength, acoustic-power, and field inclination.
Figure~\ref{gosain:fig4} shows the azimuthal variation of
various parameters along the ring marked by the white circle in
Figure~\ref{gosain:fig3}. The location of this ring corresponds
to the mid-penumbra,  with  radius of $\sim$ 8.5 arcsec. In
figure~\ref{gosain:fig4} the variation of acoustic power and
the magnetic parameters show an anti-correlation. Further, the
relation between the field strength and inclination suggests
that the stronger fields are more vertical as compared to the
weaker fields, which are more horizontal. A similar relation
was found by Schunker {\it et al.} (2005). This
anti-correlation of $B$ and $\gamma$ leads to stronger
modulation in the LOS field component $B$cos$\gamma$. This
modulation is seen in MDI-LOS magnetograms and SOT/SP
inclination maps as radial spine-like structures in penumbra,
previously identified by Lites {\it et al.} (1993).

 A spine structure and its adjacent channel (intra-spine) are
marked by two radial line segments  in the middle and bottom
panels of Figure~\ref{gosain:fig3}.  These two locations are
marked in the azimuthal profiles plotted in
Figure~\ref{gosain:fig4} by vertical lines. The spine
(intra-spine) is characterized by stronger (weaker) and
vertical (horizontal) fields. The acoustic power absorption in
the spine structure is clearly larger than in the intra-spine.
The spine corresponds a to field strength of 1400 Gauss and
inclination of 145$^\circ$ while the intra-spine corresponds to
a field strength of 1100 Gauss and inclination of 112$^\circ$.

\subsection{Radial Variation of the Parameters along Spine and Intra-spine}
In view of the azimuthal variations in magnetic and acoustic
parameters due to radial spine-like structures, discussed in
Section 3.3 above, we make a comparison of the radial variation
of these parameters along the spine and intra-spine structure.
To this end, we plot the profiles of magnetic parameters and
acoustic power along two radial line segments shown in middle
and bottom panels of Figure 5. The two line segments correspond
to  spine and intra-spine, going clockwise along the white
circle. The variation of the parameters along these magnetic
structures is shown in Figure 7. The solid (dashed) line curve
corresponds to the intra-spine (spine). It may be noted that as
compared to the mean radial profiles shown in Figure 2, the
radial extent in Figure 7 is limited to 9.6 arcsec. The reason
is that, this radial spine and intra-spine extend as coherent
structures upto 9.6 arcsec and beyond this distance other
structures from the outer penumbra start to show up, while we
are interested in the run of parameters along the spine and
intra spine.

The parameters differ in the spine and intra-spine, specially
beyond the vertical dashed line which marks the umbra--penumbra
boundary. Further, it may be noticed that, as compared to the
spine, the field strength is weaker in the intra-spine while
the transverse field is stronger. However, we see that the
power is much reduced in the spine as compared to the
intra-spine. This result suggests that the power absorption in
magnetic structures is not simply related to the transverse
component of the magnetic field, as previously suggested
\cite{Mathew2008}. This is also apparent from the behaviour of
the mean profiles of the transverse field and acoustic-power
absorption, shown in the bottom panels of Figure 2. That is,
while the profile of the transverse field on the right, and
left-hand-side of the vertical dashed line is symmetric, the
profile of the acoustic-power is not.

\section{Discussion and Conclusions}
 We have used  simultaneous MDI and \textit{Hinode} SOT/SP
observations of a sunspot, near the disk center, to study the
relationship  between the magnetic-field vector and the
\textit{p}-mode absorption. Comparing the values of the
potential transverse field derived from MDI-LOS field and the
transverse field derived from SOT/SP observations, it is found
that the former are significantly lower than the latter. This
is attributed to different sensitivities (spatial and spectral)
of the two instruments as well as different inversion methods.
Further, the observed decrease is consistent with the decrease
found by the study of the two magnetograms by Wang {\it et al.}
(2009).

 The azimuthally averaged radial profiles of field strength and inclination
show a smoothly decreasing profile, going outwards from the
umbra, in the sunspot. However, the acoustic power absorption
does not show such a smooth behaviour. At the umbra--penumbra
boundary there is a ring of enhanced \textit{p}-mode power
absorption, in agreement with the earlier study
\cite{Mathew2008}. In that study it was shown, using the
potential transverse field, that this region of enhanced
absorption is the region of strongest transverse field with
mean field inclination of about 45 degrees. However, the
availability of SOT/SP vector magnetograms allowed us to carry
out  a similar study with observed transverse field. It was
found that: \textit{i}) the maximum of potential transverse
field and observed transverse field peak at a similar radial
distance in the sunspot, \textit{ii}) this location coincides
with the region where the acoustic power is reduced, and
\textit{iii}) the profile of the acoustic power and transverse
field are not similar as one goes from the location of the peak
transverse field towards the umbra.



The azimuthal profile over the mid-penumbra, {\it i.e.} along a
ring of radius 8.5 arcsec, shows that there is significant
variation in power absorption and magnetic parameters in a
sunspot penumbra. We examine two adjacent channels one with
stronger field strength (1400 G) and more vertical field (about
145 degrees) than the other channel with slightly weaker field
strength (1100 G) and more horizontal fields (about 112
degrees). Such inhomogeneous channels show up clearly as radial
spines and intra-spines in MDI LOS maps and are much sharper in
SOT/SP inclination maps. The power absorption is greater in the
channel with larger field strength, {\it i.e.} the spine.

The presence of azimuthal inhomogeneity prompted us to look for
the radial variation of magnetic and acoustic parameters in
these channels, {\it i.e.} in the spine and intra-spine. The
field is stronger (weaker) and vertical (horizontal)  in the
spine (intra-spine). However, the transverse-field component is
stronger in the intra-spine as compared to the spine. The
acoustic-power absorption is larger in the spine, {\it i.e.}
the channel with weaker transverse field. These results suggest
that the transverse field alone may not govern the acoustic
power absorption in the magnetic structures, as suggested by
Mathew (2008) from the coincidence of maximum transverse field
and enhanced \textit{p}-mode power absorption in the mean
radial profiles of a sunspot.

In a sunspot, the magnetic-field strength and field inclination
have a strong anti-correlation \cite{Schunker2005}. So, it is
difficult to say if the difference in absorption of acoustic
signal in spines and intra-spines is due to different field
strengths or different field inclinations or combination of
both. From our study of the correlation between the acoustic
power and magnetic parameters in different regions of the
sunspot, it was found that for the entire sunspot the
anti-correlation between acoustic power and field strength is
highest (65\%), while it is lowest with the inclination (35\%).

Further, very high-resolution observations have shown that
apart from the spine and intra-spine structure of the penumbra,
the penumbral filaments possess structure in the form of dark
cores (widths $\approx$ 90 km) accompanied by lateral
brightening \cite{Scharmer2002}. Further, the radial variation
of the field inclination is quite different in the dark cores
as compared to the lateral bright component of the penumbral
filaments \cite{Langhans2005}. It would be interesting to study
the acoustic-power absorption by the small-scale features like
these dark cores in the penumbral filaments using the time
series of high-resolution dopplergrams, in future.


The leakage of \textit{p}-modes into the upper atmosphere has
been widely studied. Waves with periods around five minutes
were reported near the footpoints of coronal  loops
\cite{Berghmans1999,Moortel2002}. It was suggested that these
waves could be associated with \textit{p}-modes
\cite{Bart2005}. The formation of spicules is now believed to
be due to leakage of \textit{p}-modes which is found to be more
favourable when field lines are inclined \cite{Bart2004}. For a
given level of \textit{p}-mode excitation, the power of
acoustic waves will decrease with increase in the leakage of
the \textit{p}-modes. Why then do the intra-spines which are
more inclined field (112 degrees) show more acoustic power than
the spines which are less inclined (145 degrees)? A
simple-minded explanation might be given as follows. The
suppression of convection by magnetic fields in a highly
conducting plasma is well known, and also it is known that
convection in rolls is possible wherever the magnetic field has
large inclination to the gravity vector
\cite{Chandra61,Parker79}. Thus, the increase in power of five
minute oscillations with inclination could be due to convective
motions, which presumably excite the oscillations, and are more
vigorous in regions of large field inclination.


Another wave phenomenon seen in chromospheric images of
sunspots is that of running penumbral waves (RPW) discovered by
Zirin and Stein (1972). These are seen as bright circular
fronts originating at the umbra--penumbra boundary and moving
towards outer edge of penumbra with an interval of 300 seconds.
Recent studies suggest these waves to be slow-mode waves
propagating along the field lines \cite{Bloomfield2007}.
Further, it is suggested that the chromospheric oscillations
with five minute periods at umbra--penumbra boundary may arise
because of the field inclination becoming favourable for
\textit{p}-mode power to tunnel through acoustic cut-off
\cite{Bart2004}. The co-existence of the observed reduction in
\textit{p}-mode power and the excitation of RPWs at the
umbra--penumbra boundary suggests the possibility of a
connection between these two effects. Application of wavelet
analysis can bring out the episodal relationship between a
time-dependent variation of the modal power and the excitation
of the RPWs. This investigation will be deferred to a future
paper.

 Further, to rule out instrumental effects we plan to extend
the present study using more observations from different
instruments like (i) using the intensitygrams from Solar
Optical Telescope (SOT) onboard {\it Hinode} mission (Kosugi
\textit{et al.} 2007) and (ii) dopplergrams from Helioseismic
and Magnetic Imager (HMI) onboard Solar Dynamics Observatory
(SDO) (Scherrer and HMI Team 2002).

\begin{acks}
We thank the anonymous referee for their valuable comments and
suggestions, which helped to improve the quality of the paper.
Also, the authors thank the SOHO/MDI consortia for their data.
SOHO is a joint project by ESA and NASA.  {\it Hinode} is a
Japanese mission developed and launched by ISAS/JAXA, with NAOJ
as domestic partner and NASA and STFC (UK) as international
partners. It is operated by these agencies in co-operation with
ESA and NSC (Norway).
\end{acks}

\begin{figure}
\centerline{\includegraphics[width=1.0\textwidth,clip=]{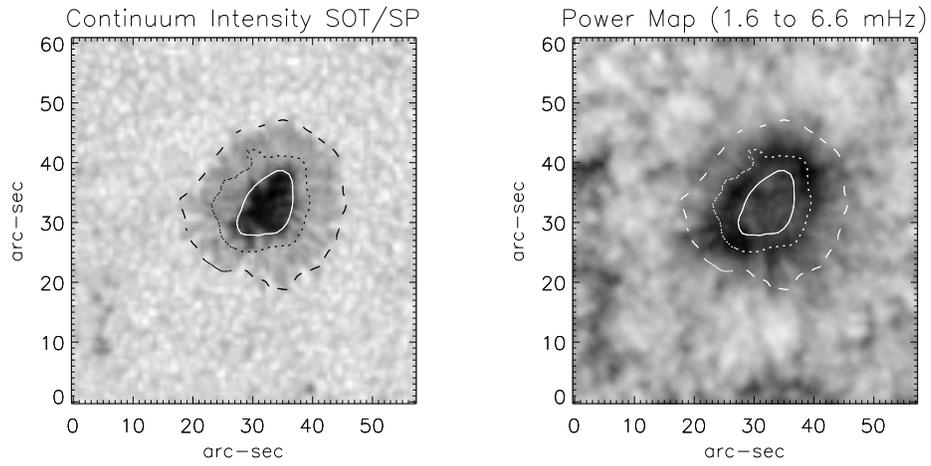}}
\caption{ The continuum intensity levels used to isolate the three regions
{\it i.e.}, umbra, umbra-penumbra boundary, and penumbra. These region bounds are shown by
solid, dotted and dashed contours over the SOT/SP continuum-intensity image (left) and acoustic-power map (right), respectively.
  }
\label{gosain:fig6}
\end{figure}

\begin{figure}
\centerline{\includegraphics[width=1.05\textwidth,clip=]{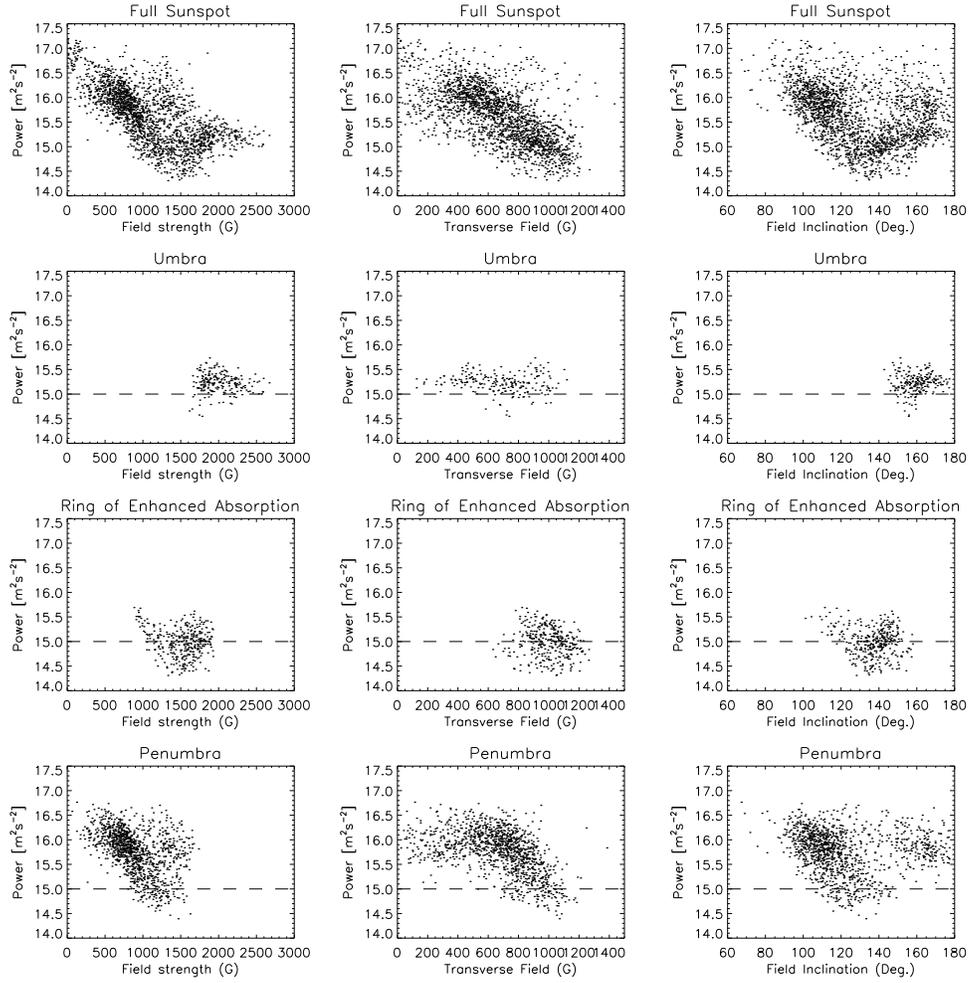}}
\caption{
The scatter between acoustic-power absorption and field
strength, transverse field, and inclination angle. The scatterplots for the full sunspot,
umbra, umbra-penumbra boundary, and penumbra are shown from top
to bottom. }
\label{gosain:fig7}
\end{figure}

\begin{figure}
\centerline{\includegraphics[width=1.05\textwidth,clip=]{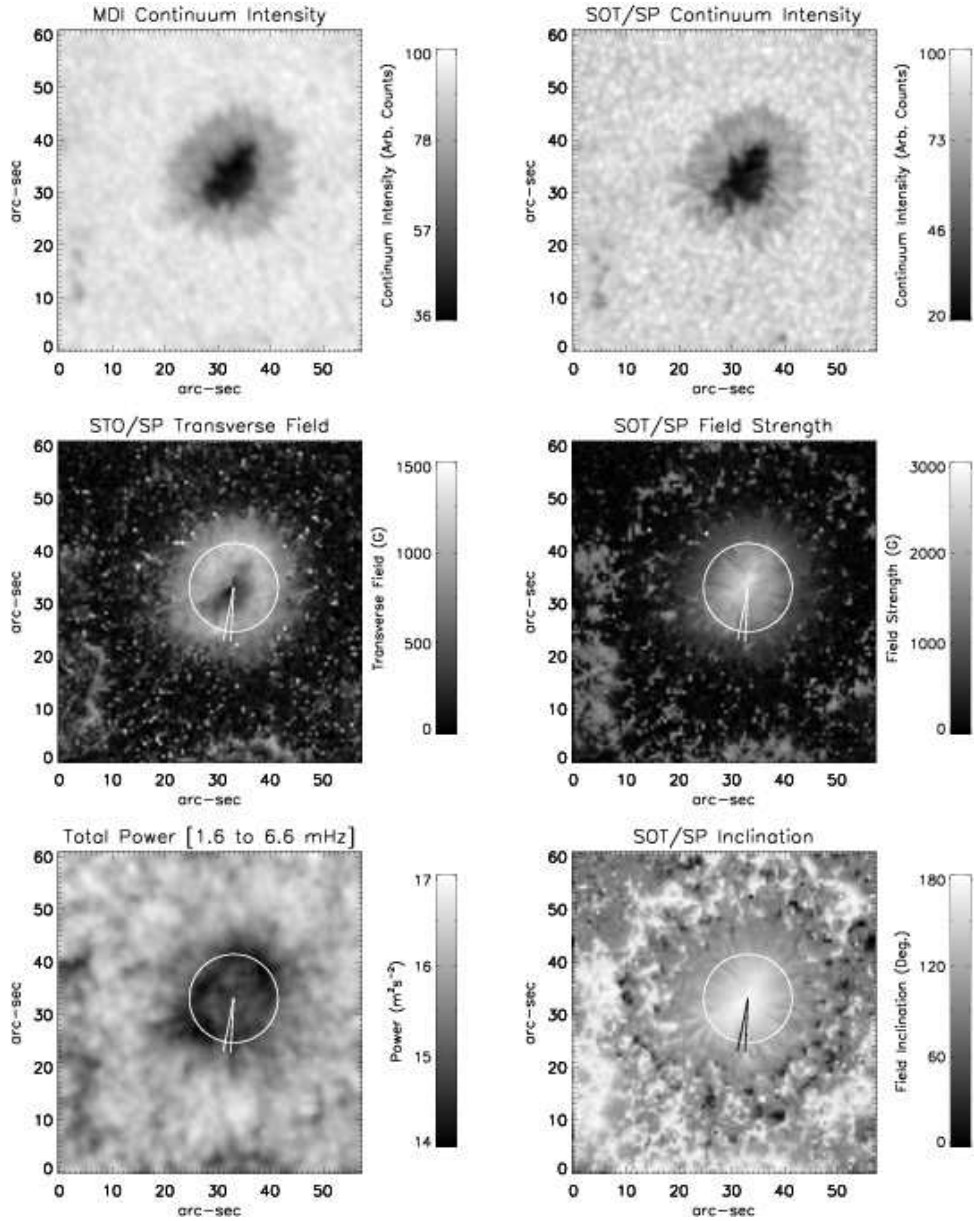}}
\caption{{\it Top row}: Continuum intensity image of the sunspot by MDI (left panel)
and SOT/SP (right panel). {\it Middle row}: transverse field (left) and
field strength (right) maps
obtained by SOT/SP observations. {\it Bottom row}: the MDI acoustic power map (left) and magnetic field
inclination (right). The white circle marks the
location for which the azimuthal profile is plotted in
Figure~\ref{gosain:fig4}. The two radial line segments in the
middle and bottom rows  mark the azimuth positions of a
spine and intra-spine (going clockwise along white circle). These positions are drawn as vertical
lines in Figure 4. The profile of the parameters along these
two line segments is shown in Figure~\ref{gosain:fig5}.}
\label{gosain:fig3}
\end{figure}

\begin{figure}
\centerline{\includegraphics[width=0.95\textwidth,clip=]{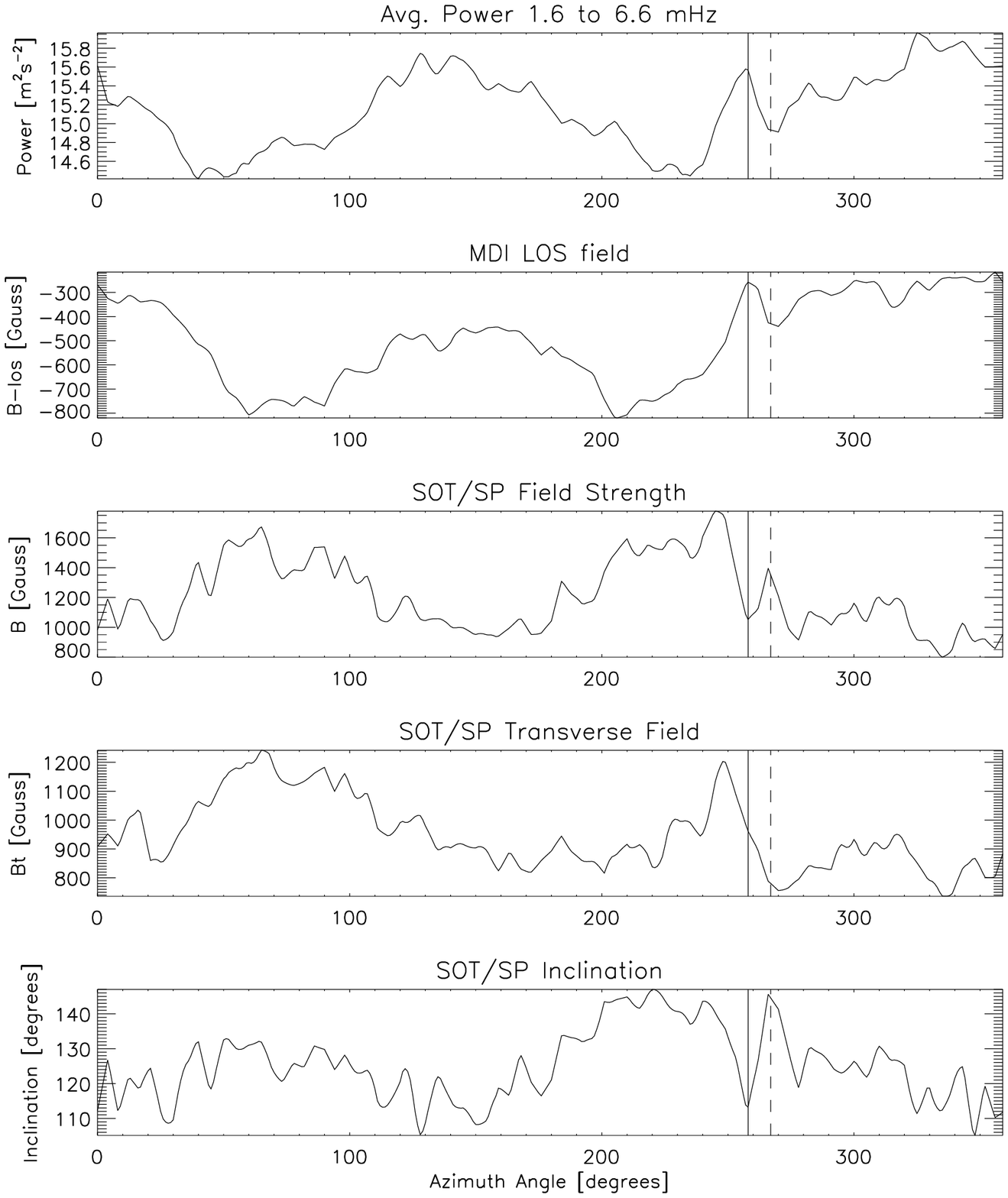}}
\caption{
The azimuthal variation of the parameters along the white
circle  of radius 8.5 arcsec in the lower panels of Figure 3. The
vertical lines, one solid and one dashed, correspond to intra-spine and spine
 structures in the penumbra, respectively. The panels from top to bottom
 represent maps of {\it i}) mean acoustic power in the band 1.6 to 6.6 mHz, {\it ii}) MDI B-los, {\it iii}) SOT/SP field strength,
{\it iv}) SOT/SP transverse field strength and {\it v}) the SOT/SP field inclination.}
\label{gosain:fig4}
\end{figure}

\begin{figure}
\centerline{\includegraphics[width=0.95\textwidth,clip=]{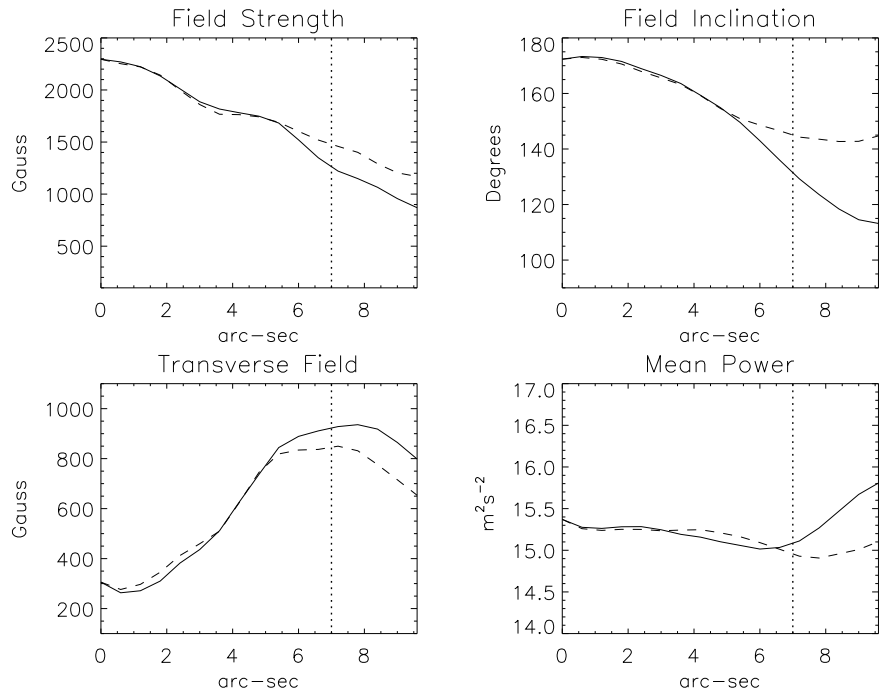}}
\caption{ The radial profiles of the parameters along the  two radial line segments in the bottom panels of Figure 3.
 The solid (dashed) line corresponds to intra-spine (spine). The vertical
 dashed line corresponds to the location of
 enhanced power absorption in the mean azimuthal profile of Figure 2.  }
\label{gosain:fig5}
\end{figure}

\end{article}

\end{document}